\documentclass[final,authoryear,5p]{elsarticle}
\usepackage{epsfig}
\usepackage{amssymb}
\usepackage[ps2pdf,%
a4paper=true,%
breaklinks=true,%
colorlinks=true,%
pdfauthor={First Author et al.},%
pdftitle={Template for manuscripts in Advances in Space Research}%
]{hyperref}
\journal{Advances in Space Research}

\newcommand{\sw}{{\it Swift}}

\newcommand{\xmm}{XMM--Newton}

\def \src {\mbox{IGR~J16328$-$4726}}
\def \ATel {The Astronomer's Telegram } 
\def \apj {ApJ }
\def \apjs {ApJS }
\def \apjl {ApJL }
\def \aj {AJ }

\def \aap {A\&A }
\def \baas {BAAS }

\def \mnras {MNRAS }
\def \pasj {PASJ }

\newcommand\ssr{Space~Sci.~Rev. }
\newcommand\farcs{\hbox{$.\!\!^{\prime\prime}$}}
\newcommand\fs{\hbox{$.\!\!^{\mbox{\small s}}$}}

\begin{document}

\begin{frontmatter}

\title{The {\it Swift} Supergiant Fast  X-ray Transients Project\tnoteref{footnote1}: \\
a review, new results, and future perspectives}
\tnotetext[footnote1]{  \href{http://www.ifc.inaf.it/sfxt/}{Project web page: http://www.ifc.inaf.it/sfxt/ }}

\author[a]{P.\ Romano\corref{cor}} 
\ead{romano@ifc.inaf.it}
\author[a]{V.\ Mangano}
\author[b]{L.\ Ducci}
\author[c]{P.\ Esposito}
\author[a]{S.\ Vercellone}
\author[d]{F.\ Bocchino}
\author[e]{D.\ N.\ Burrows}
\author[e]{J.\ A.\ Kennea} 
\author[f,g]{H.A.~Krimm} 
\author[f]{N.\ Gehrels}
\author[h]{R.\ Farinelli}
\author[h]{C.\ Ceccobello}

\address[a]{INAF-IASF Palermo,  Via U.\ La Malfa 153, I-90146 Palermo, Italy}
\cortext[cor]{Corresponding author}

\address[b]{Institut f\"ur Astronomie und Astrophysik,
         Universit\"at T\"ubingen, Sand 1, D-72076 T\"ubingen, Germany }
\address[c]{INAF-IASF Milano, Via E.\ Bassini 15,   I-20133 Milano,  Italy}
\address[d]{INAF-Osservatorio Astronomico di Palermo, Palermo, Italy}
\address[e]{Department of Astronomy and Astrophysics, Pennsylvania State 
             University, University Park, PA 16802, USA}
\address[f]{NASA/Goddard Space Flight Center, Greenbelt, MD 20771, USA}
\address[g]{Universities Space Research Association, Columbia, MD 21044-3432, USA}
\address[h]{University of Ferrara, Ferrara, Italy}

\begin{abstract}
We present a review of the Supergiant Fast X-ray Transients (SFXT) Project,  
a systematic investigation of the properties of SFXTs with a strategy that combines 
{\it Swift} monitoring programs with outburst fol\-low-up observations. 
This strategy has quickly tripled the available sets of broad-band data of SFXT outbursts,
and gathered a wealth of out-of-outburst data, 
which have led us to a broad-band spectral characterization, an assessment of the 
fraction of the time these sources spend in each phase, and their duty cycle of inactivity. 
We present some new observational results obtained through our outburst follow-ups,
as fitting examples of the exceptional capabilities of {\it Swift} in catching
bright flares and monitor them panchromatically.  
\end{abstract}

\begin{keyword}
\PACS 97.80.Jp \sep 98.70.Qy \\
X-rays: binaries \sep X-rays: individual (IGR~J16328$-$4726, IGR~J16418$-$4532, IGR~J16479$-$4514, 
XTE~J1739$-$302, IGR~J17354$-$3255, IGR~J17544$-$2619, 
AX~J1841.0$-$0536, AX~J1845.0$-$0433, IGR~J18483$-$0311)
\end{keyword}

\end{frontmatter}

\parindent=0.5 cm

\section{The {\it Swift} SFXT Project: a review \label{cospar12:intro}}

In the last decade and a half, observations by X--ray satellites
led to the discovery of a new class of high mass X--ray 
binaries (HMXBs) with supergiant companions, called
supergiant fast X-ray transients 
\citep[SFXTs; ][]{Smith2004:fast_transients,Sguera2005,Negueruela2006:ESASP604}.
The members of this class exhibit periods of enhanced 
X--ray activity with durations ranging from a few hours to a few days
and peak luminosities of $10^{36}-10^{37}$~erg~s$^{-1}$. 
A distinctive property of SFXTs is the high dynamic range,
spanning three to five orders of magnitude,
with sudden increases in luminosity from $\sim 10^{32}$~erg s$^{-1}$
up to the flare peak luminosity (e.g.\ \citealt{zand2005}).
The X-ray spectra can be described with models typically
used to fit the X-ray emission from pulsars in HMXBs
(see e.g. \citealt{Walter2007}).
We generally distinguish between confirmed and candidate SFXTs 
based on the availability of an optical classification of the companion. 
We currently have 10 confirmed and about as many candidate SFXTs. 

The accretion mechanisms responsible for the fast and high
variability of SFXTs are still poorly known and understood.
They can be divided in two categories:
those models for which the X-ray variability exclusively
depends on the properties of the geometry and inhomogeneity
of the stellar wind from the donor star
(see e.g. \citealt{zand2005}; \citealt{Sidoli2007}, 
\citealt{Negueruela2008}),
and the accretion mechanisms linking the observed high dynamic ranges 
to the properties of the compact object 
and assuming only modest variation in the density and/or velocity of the supergiant wind
(see e.g.\ \citealt{Grebenev2007}; \citealt{Bozzo2008}).

Our {\it Swift} SFXT Project 
has been performing a systematic investigation of the properties of SFXTs with a strategy that combines 
{\it Swift}\footnote{See \citet[][]{Gehrels2004} for a description of the {\it Swift} satellite.} 
monitoring programs with outburst fol\-low-up observations. 

\subsection{Long term monitoring campaigns\label{cospar12:sec:longterm}} 
\begin{figure}[t]
\begin{flushleft}
\vspace{-0.5truecm}
\hspace{-0.5truecm}
\centerline{\includegraphics[width=10cm,height=13truecm,angle=0]{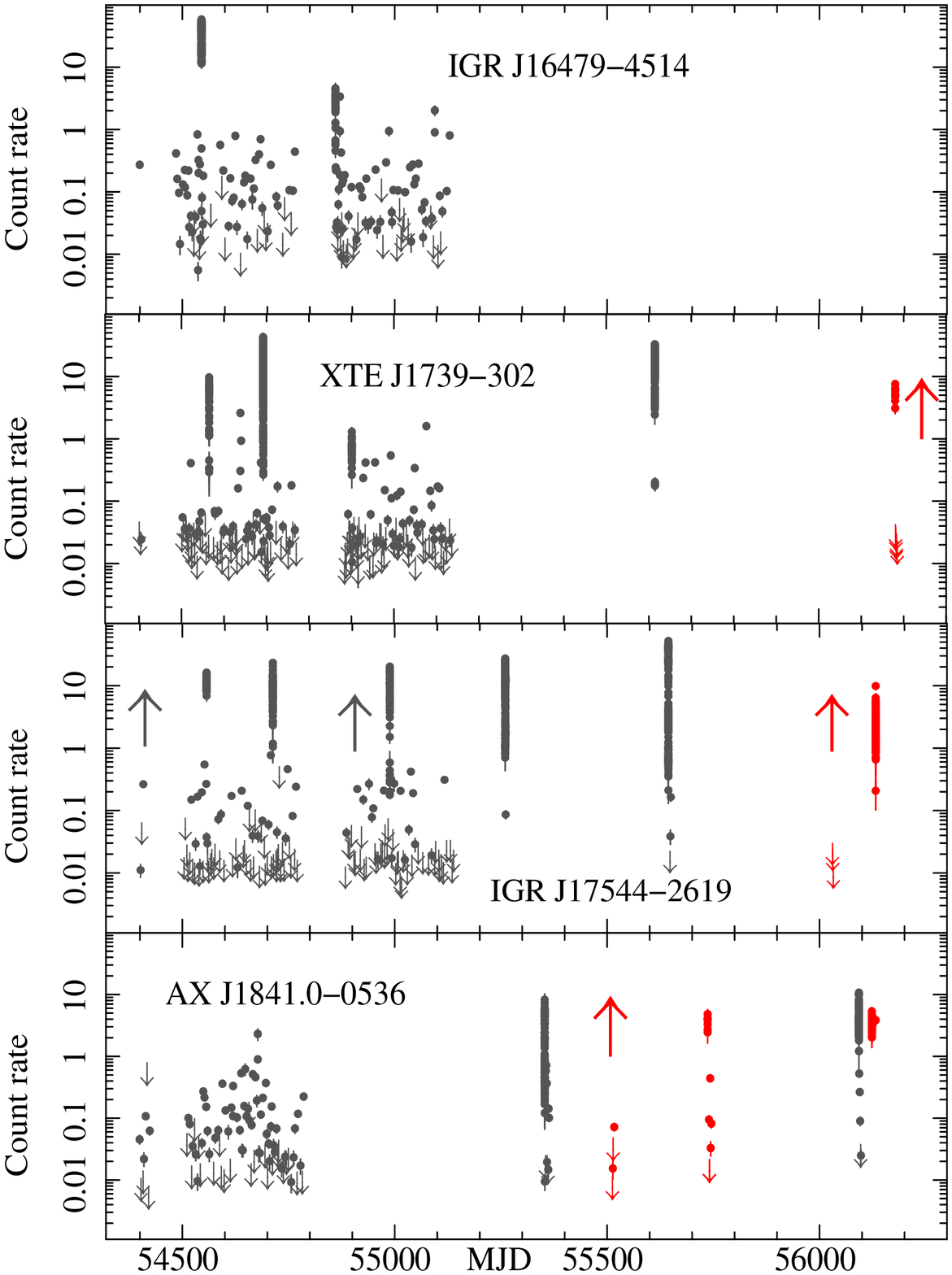}}
\end{flushleft}
\vspace{-2.0truecm}
\caption[XRT campaigns]{{\it Swift}/XRT (0.2--10\,keV) light curves
of our 2007--2009 long-term monitoring program (MJD 54399  to  55138)
and subsequent outbursts and follow-ups.  
Each point refers to the average count rate observed
during each observation performed with XRT, except for outbursts 
where the data were binned to include at least 20 counts bin$^{-1}$ to best represent the 
dynamical range. Downward-pointing arrows are 3-$\sigma$ upper limits. 
Upward pointing arrows mark bright outbursts detected 
by BAT (and not simultaneously followed by XRT),
or by MAXI (AX~J1841.0$-$0536, MJD~55507). 
Data in grey were previously published in 
\citet[][]{Romano2009:sfxts_paperV,Romano2011:sfxts_paperVI,Romano2011:sfxts_paperVII,Romano2013:MI50x_sfxts}
and \citet{Farinelli2012:sfxts_paperVIII}. Data in red are presented here for the first time. 
}
\label{cospar12:fig:campaigns}
\end{figure}

The initial monitoring sample consisted of 4 confirmed SFXTs, chosen among the 10 or so SFXTs 
known in late 2007 and it includes the two prototypes of the class XTE J1739$-$302 
and IGR~J17544$-$2619. 
The first campaigns ran from 2007 October 26 to 2009 November 3 
(MJD 54399 to 55138, shown as part of the grey data in  Fig.~\ref{cospar12:fig:campaigns}) 
with a regularly sampled pace of 2--3 observations week$^{-1}$ object$^{-1}$, each 1\,ks long,
for a total on-source exposure of 606\,ks divided in 558 observations. 
The {\it Swift}/X-ray Telescope \citep[XRT, ][]{Burrows2005:XRT}  was set in AUTO mode, 
to best exploit the  XRT automatic mode switching 
in response to changes in the observed source intensity.  
This has allowed, for the very first time, to study the long term properties of SFXTs 
with a pointed, high sensitivity X--ray telescope. 

The detailed results of these campaigns can be found in \citet[][]{Sidoli2008:sfxts_paperI} and 
\citet[][ and references therein]{Romano2009:sfxts_paperV,Romano2011:sfxts_paperVI} 
and they include the discovery, outside their outbursts, of X-ray activity in all four SFXTs, 
demonstrating that these transients accrete matter even outside their outbursts. 
This emission is highly variable, with timescales ranging from months down to the shortest timescales 
we can probe (a few hundred seconds) with a dynamic range in excess of 1 order of magnitude in all four SFXTs. 

We also assessed how long each source spends in each state \citep{Romano2011:sfxts_paperVI}
by studying the XRT count rate distributions. 
The overall dynamic range reaches then $\sim4$ orders of magnitude  when outbursts, 
that only account for 3--5\,\% of the total SFXT phase, are considered. 
The most common X-ray flux is $F_{\rm 2-10\,keV}\sim (1$--$2)\times10^{-11}$ erg cm$^{-2}$ s$^{-1}$ 
(unabsorbed), corresponding to luminosities in the order of a few 10$^{33}$--10$^{34}$~erg~s$^{-1}$, 
$\sim 100$ times lower than the bright outbursts \citep{Romano2011:sfxts_paperVI}. 
Finally, we calculated that the duty-cycle of inactivity is in the 
range $\sim 19$--$55$\,\% \citep[][]{Romano2009:sfxts_paperV,Romano2011:sfxts_paperVI}.
Therefore, differently from what was previously estimated using less sensitive instruments, 
true quiescence  (at $\sim 10^{32}$~erg s$^{-1}$)
is a relatively rare state for these transients.

\subsection{Outbursts and follow-ups\label{cospar12:sec:outburst}}  
Many observations were collected during the SFXT outbursts, as these are certainly the 
most prominent
manifestation of their activity 
\citep[][]{Romano2008:sfxts_paperII,Romano2009:sfxts_paper08408,Sidoli2009:sfxts_paperIII,Sidoli2009:sfxts_paperIV}.  
Indeed, {\it Swift} can catch and study them in broad band (0.3--150\,keV) in their early stages, 
thanks to the combination of its on-board triggering, autonomous fast slewing and panchromatic capabilities, 
thus allowing us to assess the spectroscopic and timing properties of SFXTs when they are in their brightest
phases. 
Given the shape of the SFXT spectrum (power law with an exponential cut-off), 
the large {\it Swift} energy range allows us to both constrain the 
hard-X spectral properties to compare with popular accreting neutron star models 
and to obtain a measure of the absorption. 

For this project, the Burst Alert Telescope \citep[BAT, ][]{Barthelmy2005:BAT} Team devised 
and applied to the whole sample of SFXTs and candidates,
the so called ``BAT special functions'', starting from 2008 September 25. 
They are modifications of the on-board triggering 
figure of merit that allow SFXTs to trigger the BAT, perform a slew and then observe the source 
as if it were a $\gamma$-ray burst, {\it Swift}'s natural target. This ensures that the source obtains 
not only coverage with the BAT, but also with the narrow-field instruments, XRT 
and the UV/Optical Telescope \citep[UVOT, ][]{Roming2005:UVOT}. 
Further `triggers' also came from the BAT Transient Monitor 
\citep[][]{Krimm2006_atel_BTM,Krimm2013:BATTM}\footnote{
\href{http://swift.gsfc.nasa.gov/docs/swift/results/transients/}{http://swift.gsfc.nasa.gov/docs/swift/results/transients/} }.
After the initial automated target exposure 
we generally follow the decay of the outburst for at least a week with XRT through 
target of opportunity  (ToO) observations.

\begin{figure}[H]
\begin{flushleft}
\vspace{-3truecm}
\centerline{\includegraphics[width=12cm,angle=0]{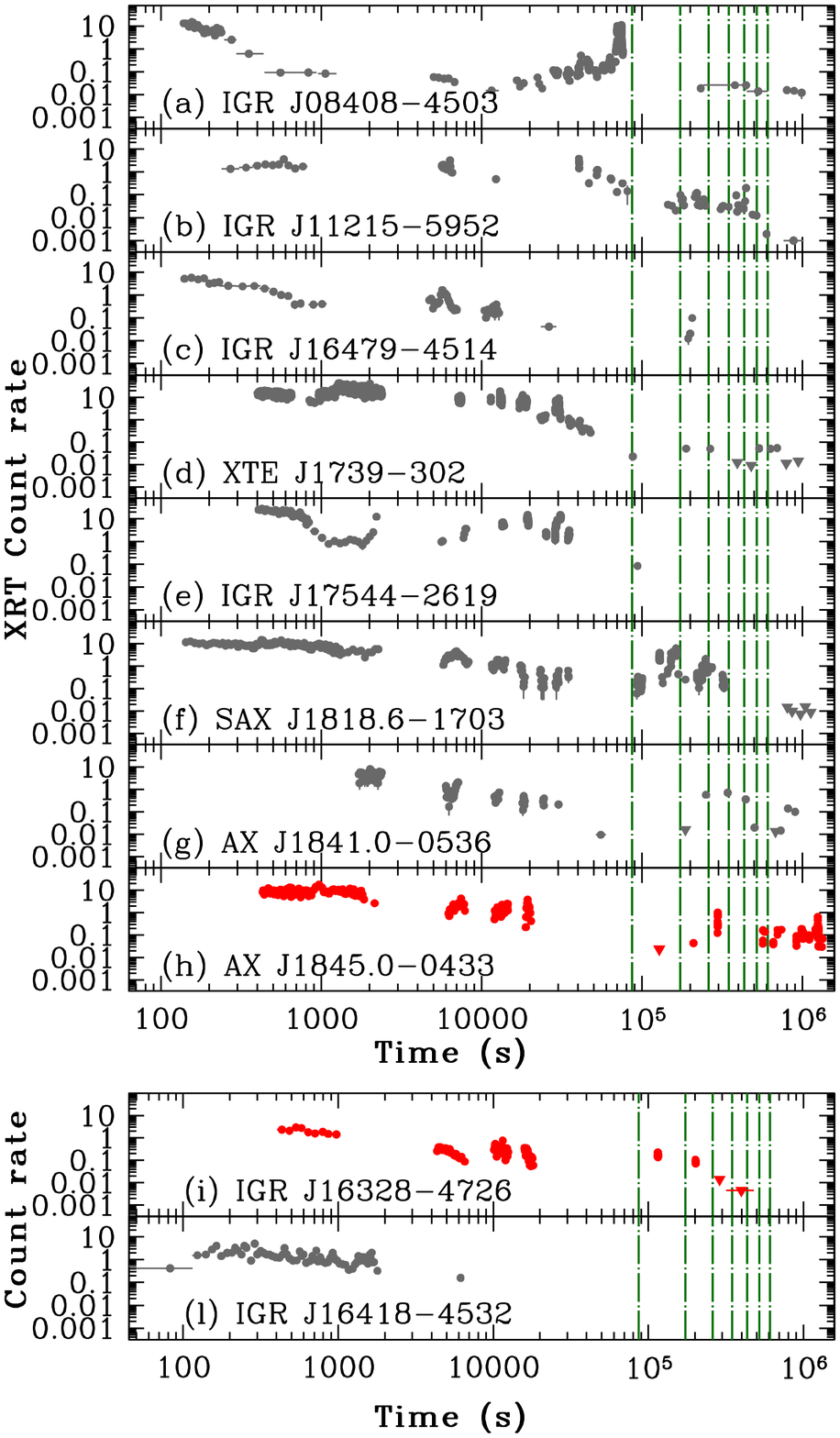}}
\end{flushleft}
\vspace{-2.0truecm}
\caption[XRT light curves]{Light 
curves of the better followed-up outbursts of SFXTs 
followed by {\it Swift}/XRT with starting times
referred to their respective BAT triggers
(IGR~J11215$-$5952 
did not trigger the BAT, so it is referred to MJD 54139.94; 
similarly, IGR~J16418$-$4532 is referred to the start of the observation).
Points denote detections, triangles 3$\sigma$ upper limits. 
Red data points (panels h and i) refer to observations presented here 
for the first time, while grey points to data presented elsewhere. 
Vertical dashed lines mark time intervals equal to 1 day, up to a week. 
References: 
IGR~J08408--4503 \citep[2008-07-05, ][panel a]{Romano2009:sfxts_paper08408};
IGR~J11215$-$5952 \citep[2007-02-09, ][panel b]{Romano2007}; 
IGR~J16479$-$4514 \citep[2005-08-30, ][panel c]{Sidoli2008:sfxts_paperI}; 
XTE~J1739$-$302 \citep[2008-08-13, ][panel d]{Sidoli2009:sfxts_paperIV}; 
IGR~J17544$-$2619 \citep[2010-03-04, ][panel e]{Romano2011:sfxts_paperVII}; 
SAX~J1818.6$-$1703 \citep[2009-05-06, ][panel f]{Sidoli2009:sfxts_sax1818};  
AX~J1841.0$-$0536 \citep[2010-06-05, ][panel g]{Romano2011:sfxts_paperVII}; 
AX~J1845.0$-$0433 (2012-05-05, this work, panel h); 
candidate SFXT IGR~J16328--4726 (2009-06-10, this work, panel i); 
candidate SFXT IGR~J16418$-$4532 \citep[2011-02-18, ][panel l]{Romano2012:sfxts_16418}.
}
\label{cospar12:fig:best_sfxts}
\end{figure}

Our strategy has several advantages. 
\begin{enumerate}
\item  It quickly tripled the available sets of broad-band data of SFXT outbursts
\citep[e.g.\ ][]{Sidoli2009:sfxts_sax1818,Romano2011:sfxts_paperVI,Romano2011:sfxts_paperVII,Romano2013:MI50x_sfxts,Farinelli2012:sfxts_paperVIII,Mangano2012:Gamma12}. 
From launch to the end of 2012, {\it Swift} has detected 42 outbursts, for a total of 45 
on-board triggers (5 from candidates), 
24 (2) with broad band coverage, 22 of which have been detected thanks to the BAT special functions being applied. 
Fig.~\ref{cospar12:fig:campaigns} (red points) shows some of the most recent outbursts of 
the 4 monitored SFXTs after the long term monitoring described in Section~\ref{cospar12:sec:longterm} ended
in 2009 November.

\item  The unique \sw\ characteristic of an automated, fast repointing to potentially 
any target that triggers the BAT, has allowed the arcsecond localization  for several SFXTs and candidates 
whose coordinates were only known to the arcminute level, or to improve upon previously known coordinates,
and consequently help in associating with an optical counterpart. 
Such is the case for \src, that we localized when it first went into outburst  \citep{Grupe2009:16328-4726}, 
and AX~J1845.0$-$0433, whose position we now refine with respect to a previously known {\it XMM-Newton} one.
Both cases are presented in detail below.  

\item We have now observed outbursts from most confirmed SFXTs 
and followed them with XRT 
for days after the outburst (by reaching a few tenths of a $\mu$Crab in 10\,ks), 
well after it had become undetectable with monitoring instruments with lower 
sensitivity (especially in crowded fields) such as the All-Sky Monitor on board 
RXTE (\citealt[][10 mCrab in the 1.5--12\,keV band at 2$\sigma$ level for 1 day average)]{Levine1996:RXTE_ASM}
or MAXI (\citealt[][15\,mCrab in the 0.5--20\,keV band at 5$\sigma$ level for 1 day average)]{MAXI:Matsuoka2009PASJ}. 
The simplest piece of information we can derive from the X-ray light curves
is the dynamic range, which is a discriminant \citep[][]{Negueruela2006:ESASP604,Walter2006} 
between outbursts of classical supergiant HMXB (sgHMXB, $\lesssim 20$) and SFXT ($\gtrsim100$). 
Fig~\ref{cospar12:fig:best_sfxts} shows the best examples of XRT outburst light curves, 
and highlights what we have discovered as the common X--ray characteristics of this class: 
outburst lengths well in excess of hours (sometimes lasting several days), with a multiple-peaked structure,  
and a dynamic range (only including bright outbursts) up to $\sim3$ orders of magnitude. 
\end{enumerate}

\subsection{A Physical model for SFXT broad-band spectra\label{cospar12:sec:compmag}.} 

As part of our ongoing effort to understand the physical mechanisms responsible for
the bright outbursts, we have recently developed a physical model, {\tt compmag} 
in {\tt XSPEC} 
\citep[][]{Farinelli2012:compmag}, 
which includes thermal and bulk Comptonization for cylindrical accretion onto a 
magnetized neutron star. 
 A full description of the algorithm can be found in \citet[][]{Farinelli2012:compmag};  
here we summarize the main characteristics and the results of the first
applications.

\clearpage

The {\tt compmag}  model considers a blackbody spectrum of photons 
as seed for Comptonization of the plasma. 
The free parameters are temperature $kT_{\rm bb}$ of the blackbody seed photons, 
the electron temperature $kT_{\rm e}$,  and vertical optical depth $\tau$ of the 
Comptonization plasma, 
the radius of the accretion column $r_{\rm 0}$ 
and the albedo $A$ at the surface of the neutron star. 
The velocity field of the accreting matter can 
either be increasing towards the NS surface, in which case further free parameters are 
the terminal velocity $\beta_0$ at the NS surface, and the index of the law 
$\beta(z) \propto z^{-\eta}$, 
or it may be described by an approximately decelerating profile, so that the velocity law
is given by $\beta(\tau) \propto -\tau$. 

The {\tt compmag} model was first applied to 
the SFXT class prototypes XTE J1739$-$302 and IGR~J17544 $-$2619 
\citep[][]{Farinelli2012:sfxts_paperVIII}.  
We note that, in general, it is not advisable to keep all the above mentioned parameters free, 
when trying to constrain them. 
We therefore assumed an increasing velocity of the infalling material towards the NS surface 
and set $\eta=0.5$ and $\beta_0= 0.2$ or 0.05. 
We also set $r_{\rm 0}= 0.25 $  (in units of the NS Schwarzschild radius) and $A=1$.
Under these assumptions, we found that the XRT+BAT spectra of the 
2011 February 22  and 2011 March 24 outbursts of XTE J1739$-$302 and IGR~J17544 $-$2619,
respectively, can be fit well with a single unsaturated Comptonization model such as 
 {\tt compmag}. In particular, we showed that the electron density 
in the region of the X-ray spectral formation, inferred from the best-fitting parameters 
of the {\tt compmag} model, is a about 3 orders of magnitude higher than expected 
from the continuity equation at the magnetospheric radius, and we proposed that this might be 
explained in terms of a bow shock at the NS magnetosphere. 
This effect is currently under investigation via 3D magnetohydrodynamical simulations.

\begin{figure}[t]
\begin{flushleft}
\vspace{-1.5truecm}

\hspace{-0.5truecm}
\centerline{\includegraphics[width=10.5cm,angle=0]{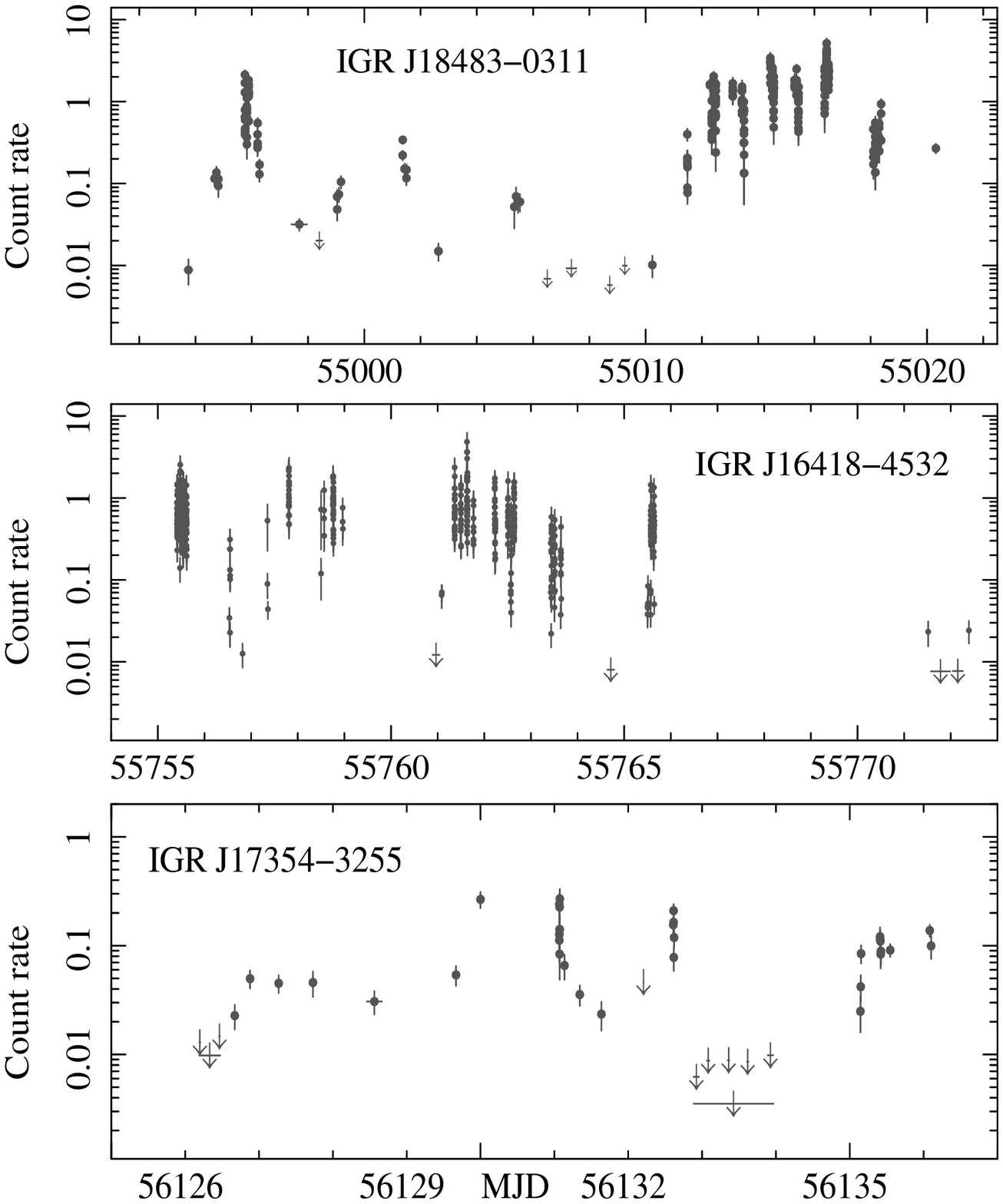}}
\end{flushleft}
\vspace{-3truecm}
\caption[XRT campaigns]{{\it Swift}/XRT light curves of monitoring programs along 
one or more orbital periods.  
{\bf Top:} IGR~J18483$-$0311, with orbital period $P_{\rm orb}=12.545$\,d \citep{Levine2011}, 
observed on 2009 June 11 to July 9, shown at a binning of at least 20 counts bin$^{-1}$ 
\citep[adapted from][]{Romano2010:sfxts_18483}.
{\bf Middle:} IGR~J16418$-$4532, with  $P_{\rm orb}=3.73886$\,d \citep{Levine2011}, 
observed on 2011 July 13--30, shown at a binning with a S/N$=3$
\citep[adapted from][]{Romano2012:sfxts_16418}. 
{\bf Bottom:} IGR~J17354$-$3255, with  $P_{\rm orb}=8.4474$\,d \citep{Sguera2011:17354}, 
observed on 2012 July 18--28,  shown at a binning of at least 20 counts bin$^{-1}$ 
\citep[adapted from][]{Ducci2013:sfxts_17354}. 
}
\label{cospar12:fig:campaigns2}
\end{figure}

\subsection{Further monitoring campaigns\label{cospar12:sec:monit2}.} 
Further monitoring campaigns featured observations \,
along one or more orbital periods of an SFXT, to study the effects of orbital parameters on the flare distributions.
Such is the case of IGR~J18483\-$-$0311, observed between 2009 June 11 and July 9 
\citep[23 daily observations, $\sim2$\,ks each, spread over 28\,d 
for a total of 44\,ks; ][see  Fig.~\ref{cospar12:fig:campaigns2} top]{Romano2010:sfxts_18483}
with orbital period $P_{\rm orb}=18.545$\,d \citep{Levine2011}, 
the candidate SFXT     IGR J16418\-$-$4532, 
observed on 2011 July 13--30 \citep[12 observations, $\sim2$--5\,ks each, spread 
over 18\,d for a total of 39\,ks; ][see Fig.~\ref{cospar12:fig:campaigns2} middle]{Romano2012:sfxts_16418},
with  $P_{\rm orb}=3.73886$\,d \citep{Levine2011},
and the candidate SFXT  IGR J17354\-$-$3255, 
observed on 2012 July 18--28 \citep[22 observations, $~\sim1$\,ks each, spread 
over 11\,d for a total of 24\,ks;  ][see Fig.~\ref{cospar12:fig:campaigns2} bottom]{Ducci2013:sfxts_17354}, 
with  $P_{\rm orb}=8.4474$\,d \citep{Sguera2011:17354}. 
These unique datasets allowed us to constrain in these objects the different mechanisms 
proposed to explain their nature. In particular, we applied the clumpy wind model for blue supergiants 
by \citet[][]{Ducci2009} to the observed X-ray light curve. 
By assuming for IGR~J18483\-$-$0311 an eccentricity of $e = 0.4$ 
and for IGR~J16418\-$-$4532 a circular orbit, 
we could explain their X-ray emission in terms of the accretion from a spherically 
symmetric clumpy wind, composed of clumps with different masses, 
ranging from $10^{18}$ to $5\times10^{21}$\,g for IGR J1848\-$-$0311, 
and from $\sim5\times10^{16}$~g to $10^{21}$~g for IGR J16418\-$-$4532. 
The soft X-ray light curve of  IGR~J17354$-$3255 shows a moderate orbital 
modulation and a dip whose nature we investigated by comparing the X-ray light 
curve with the prediction of the Bondi-Hoyle-Lyttleton accretion theory, 
assuming both spherical and non spherical symmetry of the outflow from the donor star. 
We found that dip cannot be explained with the X-ray orbital modulation. 
We propose that an eclipse or the onset of a gated mechanism are the most likely 
explanations for the observed light curve.

In the following we present some recent observational results obtained through 
our outburst follow-ups. 
As fitting examples of the exceptional capabilities of {\it Swift} in catching
bright flares and monitor them panchromatically, we report on the 
datasets of the candidate SFXT  IGR J16328\-$-$4726 and the 
confirmed SFXT AX J1845.0\-$-$0433, 
that went into outburst on 2009 June 10 \citep{Grupe2009:16328-4726}
and 2012 May 05 \citep{Romano2012:atel4095}, respectively.

         \section{Data reduction\label{cospar12:dataredu}}

The new XRT data were processed with standard procedures 
({\tt xrtpipeline} v0.12.6), filtering and screening criteria, 
within {\tt FTOOLS} in the {\tt HEASOFT} package (v.6.12). 
For the monitoring campaigns we generally considered photon-counting (PC) data only,
and selected event grades 0--12 (\citealt{Burrows2005:XRT}); the outburst light 
curves generally start off with windowed timing (WT) data, instead, 
from which we selected event grades 0--2. 
Source events were accumulated within an annular/circular region 
(depending on whether pile-up correction was required or not, respectively) 
with an outer radius of 20--30 pixels, depending on source brightness; 
background events were accumulated from source-free regions. 
For our spectral analysis, we extracted events in the same regions as 
those adopted for the light curve creation; 
ancillary response files were generated with {\tt xrtmkarf},
to account for different extraction regions, vignetting, and PSF corrections. 

The BAT data of the outbursts were analysed using the standard BAT analysis 
software within {\tt FTOOLS}. 
Mask-tagged BAT light curves were created in the standard energy bands 
and rebinned to either achieve a signal-to-noise ratio (S/N) of at least 5 
or a 50-s integration time. 
Survey data products, in the form of Detector Plane
Histograms (DPH), were analysed with the
standard {\tt batsurvey} software. 
BAT mask-weighted spectra were extracted during several time intervals;
an energy-dependent systematic error vector was applied and 
response matrices were generated with {\tt batdrmgen}.   

We derive astrometrically corrected X-ray positions by  
using the XRT-UVOT alignment and matching to the 
U\-SNO\--B1 catalog \citep[][]{Monet2003:USNO_mn} 
as described in 
\citet{Goad2007:xrtuvotpostions} and \citet{Evans2009:xrtgrb_mn}\footnote{ 
\href{http://www.swift.ac.uk//user_objects/}{http://www.swift.ac.uk//user\_objects/}}. 

Fig.~\ref{cospar12:fig:campaigns} shows all available data on the 4 SFXT
monitored in 2007--2009 in the 0.2--10\,keV band. 
While some were previously published in 
\citet[][]{Romano2009:sfxts_paperV,Romano2011:sfxts_paperVI,Romano2011:sfxts_paperVII,Romano2013:MI50x_sfxts}
and \citet{Farinelli2012:sfxts_paperVIII}, 
the data in red are presented here for the first time, and are updated to the end of 2012. 

Fig.~\ref{cospar12:fig:best_sfxts} shows the light curves of the most representative outbursts of 
confirmed SFXTs (panels a--h) and candidate SFXTs (panels i--l), that we 
followed with {\it Swift}/XRT for a few days after the BAT trigger through ToO observations. 
Red data points, 
the XRT light curves of outbursts the confirmed SFXT AX~J1845.0-0433 and candidate SFXT IGR~J16328$-$4726 
(panels h and i),  refer to observations presented here 
for the first time, while grey points indicate data presented elsewhere. 

         \section{IGR~J16328$-$4726\label{cospar12:16328}}

\begin{figure}[t]
\begin{center}
\vspace{-0.7truecm}
\centerline{\includegraphics*[angle=0,width=8.5cm]{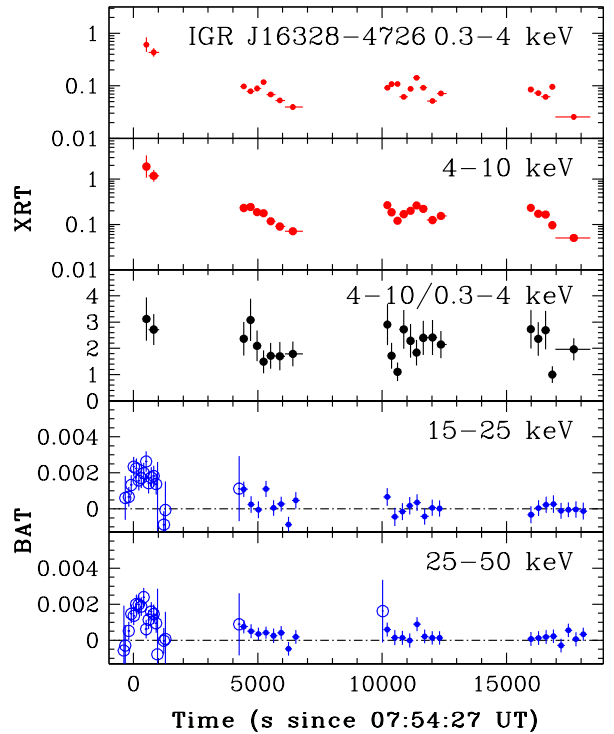}}
\end{center}
\vspace{-1.5truecm}
\caption{2009 June 10 outburst of \src: XRT and BAT light curves 
of the first day of data, in units of count s$^{-1}$ and count s$^{-1}$ detector$^{-1}$, respectively. 
The empty (blue) circles correspond to BAT in event mode data,  
filled  (blue) circles to BAT in survey mode data.  
}
\label{cospar12:fig:16328lcv_xrtbatbands}
\end{figure}

IGR~J16328$-$4726 \citep{Bird2010:igr4cat,Baumgartner2010:BAT58mos,Cusumano2010:batsur_III} 
is considered a candidate SFXT, 
based on its history of hard X-ray activity characterized by short flares 
lasting up to a few hours \citep{Fiocchi2010:16328-4726}, and a lack of confirmed counterpart, 
with an 
orbital period of $10.076\pm0.003$\,d \citep{Corbet2010:16328-4726}. 

Following the BAT trigger of \src\ on 2009 June 10 at 07:54:27  UT 
\citep[image trigger=354542,][]{Grupe2009:16328-4726}, 
\sw\ executed an immediate slew, so 
XRT data are available which are simultaneous with the BAT data. 

Using 7818\,s of XRT Photon Counting mode data and 18 UVOT images, 
we find an astrometrically corrected X-ray position:  
    RA(J$2000)$ = $16^{\rm h} 32^{\rm m} 37\fs87$, 
    Dec(J$2000)=-47^{\circ} 23^{\prime} 41\farcs2$ 
with an uncertainty of 1\farcs4 (90\,\% c.l.). 
The XRT position is $0\farcs51$ from 2MASS~J16323791$-$4723409,
the most likely optical counterpart.
This is the best currently available X--ray position for this source.

Fig.~\ref{cospar12:fig:16328lcv_xrtbatbands} shows the 0.3--4\,keV and 4--10\,keV 
light curves and the 4--10/0.3--4 hardness ratio of the 2009 June 10 outburst
during the initial bright phase.  
The remainder of the XRT data (see Fig.~\ref{cospar12:fig:best_sfxts} panel i) 
show that the count rate decreases from about 3 counts s$^{-1}$ (first orbit) 
by a factor of 10 during the first 20\,ks of observations. 
During the following two days, the source was observed at about 
0.1 count s$^{-1}$, thus providing us with a dynamic range of $\approx 40$. 
The last two observations only yield 3$\sigma$ upper limits at $1.4 \times 10^{-2}$ 
counts s$^{-1}$ and $4.5\times10^{-3}$ counts s$^{-1}$, so that the overall dynamic 
range reaches at least 650. 
While this dynamic range is lower than the 4--5 orders of magnitude observed in ``classical''
\citep{Chaty2010arXiv1012.2318C} SFXTs (such as IGR~J17544$-$2619 and XTE~J1739$-$302), 
it nonetheless places \src\ in the typical range for intermediate SFXTs 
(e.g., IGR J18483\-$-$0311 and IGR J16418\-$-$4532, \citealt[][]{Romano2010:sfxts_18483,Romano2012:sfxts_16418}). 
Fig.~\ref{cospar12:fig:best_sfxts} (panel i) shows several flares, as also typical
of the SFXT behaviour after the initial bright flare. 
We note that \citet[][]{Bozzo2012:HMXBs} reported on a 22\,ks \xmm\  
observation performed on 2011 February 20, in which they observed the source in 
a flux state slightly fainter that the one observed during the outbursts
(unabsorbed $F_{\rm 2-10\,keV}=1.7\times10^{-11}$  erg cm$^{-2}$ s$^{-1}$) also 
characterized by luminosity variations by a factor of 10.

 \begin{table}
 \begin{center}
 \caption{Spectral fits of simultaneous XRT and BAT data of \src. POW: absorbed powerlaw. CPL: cutoff powerlaw, energy cutoff E$_{\rm c}$ (keV). 
    HCT: absorbed powerlaw, high energy cutoff E$_{\rm c}$ (keV), e-folding energy E$_{\rm f}$ (keV).
Absorbing column density is in units of $10^{22}$ cm$^{-2}$. }
 \label{cospar12:tab:16328broadspec}
 \begin{tabular}{lrrrrr}
 \hline
  \noalign{\smallskip}
Model   & $N_{\rm H}$   &$\Gamma$  &$E_{\rm c}$  &$E_{\rm f}$  &$\chi^{2}_{\nu}$/dof \\
  \noalign{\smallskip}
 \hline
 \noalign{\smallskip} 
POW &$12.3_{-1.9}^{+2.3}$ &$1.66_{-0.32}^{+0.32}$ &               &                       &$1.38/51$ \\
CPL &$8.4_{-1.9}^{+2.2}$ &$0.60_{-0.48}^{+0.50}$ &$19_{-7}^{+15}$ &                        & $1.09/50$\\ 
HCT &$8.8_{-1.8}^{+2.1}$ &$0.93_{-1.79}^{+2.12}$ &$<28$ &$20_{-8}^{+16}$          & $1.09/49$ \\ 
 \noalign{\smallskip}
  \hline
  \end{tabular}
  \end{center}
  \end{table}

Fig.~\ref{cospar12:fig:16328lcv_xrtbatbands}  also shows 
the BAT light curves during the brightest part of the 2009 June 10 outburst 
in the 15--25\,keV and 25--50\,keV energy bands. 
The source is not detected above $\sim 70$\,keV. 

We extracted the mean BAT mask-weighted spectrum    
during the first orbit of data of the 2009 June 10 outburst 
and fit it in the 15--70\,keV energy band with a simple
power law. We obtained 
$\Gamma_{\rm BAT}=2.6\pm0.4$ ($\chi^2_{\nu}=1.235$, 37 degrees of freedom, dof), and 
a 20--50\,keV flux of  $7.1\times10^{-10}$ erg cm$^{-2}$ s$^{-1}$.  
We extracted the mean XRT spectrum of the brightest X-ray emission 
(obs.\ 00354542000, first orbit $T+$404--1036\,s, 632\,s net exposure,
hereafter 
`000orb1'), and 
fit it in the 0.5--10\,keV energy band with an absorbed power-law model 
by adopting \citet{Cash1979} statistics.  
We modelled the absorption ({\tt tbabs} in {\tt xspec}) with the abundance 
set to the values given by \citet{Wilms2000}. 
We obtained a column of $N_{\rm H}=9_{-3}^{+4}\times10^{22}$\,cm$^{-2}$
(in excess of the Galactic one, 
$N_{\rm H}^{\rm Gal}=1.54\times 10^{22}$\,cm$^{-2}$; \citealt{LABS}), 
a photon index $\Gamma=0.65_{-0.64}^{+0.69}$ 
(Cash statistics 618.3, 76\,\% of $10^4$ realizations with 
statistics in excess of Cstat), and 
2--10\,keV observed and unabsorbed fluxes of  
$F_{\rm 2-10\,keV}^{\rm obs}=2.9\times10^{-10}$\,erg\,cm$^{-2}$ s$^{-1}$  
and $F_{\rm 2-10\,keV}$ $=  4.2\times10^{-10}$\,erg\,cm$^{-2}$ s$^{-1}$, respectively. 
Assuming a distance of 10\,kpc 
for the luminosity calculation, we obtained $L_{\rm 2-10\,keV}=5\times10^{36}$ erg\,s$^{-1}$. 

\begin{figure}[t]
\begin{center}
\vspace{+0.4truecm}
\centerline{\includegraphics[angle=270,width=8.5cm]{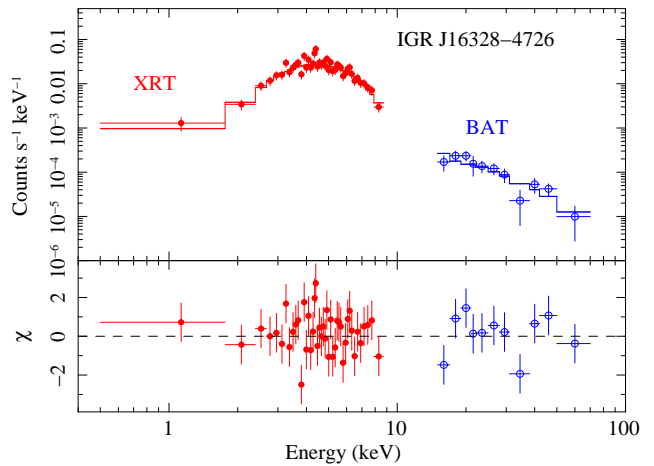}}
\end{center}
\caption{Broad-band spectroscopy of the 2009 June 10 outburst of \src.    
		Top: simultaneous XRT/PC (filled red circles) and BAT (empty blue circles) data 
			fit with the {\tt cutoffpl} model. 
		Bottom: the residuals of the fit (in units of standard deviations). }
\label{cospar12:fig:16328meanspec}
\end{figure}

 Similarly, the mean XRT spectrum, extracted from the whole first observation 
(obs.\ 00354542000, $T+$404--18,\,375\,s, 8181\,s net exposure, hereon `000') and 
rebinned with a minimum of 20 counts bin$^{-1}$ 
to allow $\chi^2$ fitting, yielded  
$N_{\rm H}=9.0_{-1.8}^{+2.2}\times 10^{22}$ cm$^{-2}$, 
and   $\Gamma=0.96_{-0.37}^{+0.41}$ 
($\chi^2_{\nu}=1.07$, 41 dof). 
The 2--10\,keV observed and unabsorbed fluxes are   
$F_{\rm 2-10\,keV}^{\rm obs}=(5.6_{-2.7}^{+0.3})\times10^{-11}$ erg cm$^{-2}$ s$^{-1}$ 
and $F_{\rm 2-10\,keV}=(8.3_{-1.6}^{+0.4})\times10^{-11}$ erg cm$^{-2}$ s$^{-1}$,
respectively; this corresponds to a luminosity 
$L_{\rm 2-10\,keV} = 1\times10^{36}$ erg s$^{-1}$.

In order to perform broad-band spectroscopy of the 2009 June 10 outburst,  
we matched the BAT spectrum with the `000' XRT one.   
We note that the chosen spectra are not strictly simultaneous, as they 
span the first and the first 4 orbits of data, respectively. 
The XRT `000' spectrum, however, offers the advantage of higher S/N than 
the strictly simultaneous `000orb1', and a consistent photon index, 
despite a decrease in average flux by a factor of $\sim 5$. 
Therefore, constant factors were included in the fitting to allow for 
both normalization uncertainties between the two instruments (generally 
constrained within 10\,\%) and normalization differences due to the non 
strict simultaneity of the XRT and the BAT data. 
We considered models that are generally used to describe the X--ray emission 
from accreting pulsars in HMXBs. i.e., a simple absorbed power-law, 
an absorbed power-law with an exponential cut-off ({\tt cutoffpl} in {\tt xspec}), 
and an absorbed power-law with a high energy cut-off ({\tt highecut}). 
We performed joint fits in the 0.5--10\,keV, 15--70\,keV 
energy bands for XRT and BAT, respectively. 
Our results are reported in Table~\ref{cospar12:tab:16328broadspec}. 
The simple power-law model yields a clearly unsatisfactory fit. 
The absorbed power-law with a high energy cut-off yields a poorly constrained
value of the cut-off energy. The absorbed power-law with an exponential cut-off,
on the other hand, yields a very good fit 
($\chi^{2}_{\nu}$/dof$=1.09/50$; see Fig.~\ref{cospar12:fig:16328meanspec}),
with $N_{\rm H}=(8\pm2)\times 10^{22}$ cm$^{-2}$,  $\Gamma=0.60_{-0.48}^{+0.50}$,
and a high-energy cut-off at $E_{\rm c}=19_{-7}^{+15}$ keV.

         \section{AX~J1845.0$-$0436\label{cospar12:18450}}

 \begin{table}
 \begin{center}
 \caption{Spectral fits of simultaneous XRT and BAT data of AX~J1845.0$-$0433. 
POW: absorbed powerlaw. CPL: cutoff powerlaw, energy cutoff E$_{\rm c}$ (keV). 
    HCT: absorbed powerlaw, high energy cutoff E$_{\rm c}$ (keV), e-folding energy E$_{\rm f}$ (keV).
Absorbing column density is in units of $10^{22}$ cm$^{-2}$. 
} 
\vspace{9pt}
 \label{cospar12:tab:18450broadspec}
 \begin{tabular}{lrrrrr}
 \hline
  \noalign{\smallskip}
 Model   & $N_{\rm H}$   &$\Gamma$  &$E_{\rm c}$  &$E_{\rm f}$  &$\chi^{2}_{\nu}$/dof \\
  \noalign{\smallskip}
 \hline
 \noalign{\smallskip} 
POW & $2.0_{-0.4}^{+0.5}$ & $1.20_{-0.25}^{+0.26}$   &                      &                         & $1.24/78$ \\ 
CPL  & $1.3_{-0.3}^{+0.4}$  & $0.56_{-0.25}^{+0.26}$  & $17_{-5}^{+8}$ &                        & $0.79/77$\\  
HCT & $1.2_{-0.3}^{+0.4}$  & $0.60_{-0.21}^{+0.31}$  &$<19$  & $18_{-5}^{+8}$   & $0.78/76$ \\  
 \noalign{\smallskip}
  \hline
  \end{tabular}
  \end{center}
  \end{table} 

AX~J1845.0$-$0433/IGR J18450$-$0435 
\citep{Bird2010:igr4cat,Baumgartner2010:BAT58mos,Cusumano2010:batsur_III} 
was discovered in ASCA data \citep[][]{Yamauchi1995:1845} as a source 
variable on timescales of tens of minutes and classified as a SFXT 
\citep[][]{Sguera2007:18450} with an O9.5I companion \citep[][]{Coe1996:18450,Zurita2009:1845}. 

AX~J1845.0$-$0433 triggered the BAT on 2012 May 05 at 01:44:39 UT\footnote{
Previously, {\it Swift} caught a flare from this source on 2005 November 4 
and 2009 June 28 \citep{Romano2009:atel2102}. 
The historical light curve from the BAT hard X-ray transient monitor can be found at 
\href{http://swift.gsfc.nasa.gov/docs/swift/results/transients/weak/IGRJ18450-0435/}{http://swift.gsfc.nasa.gov/docs/swift/results/transients/weak/}. 
}   \citep[image trigger=521567,][]{Romano2012:atel4095}. 
\sw\ slewed to the target immediately, so that the XRT began observing the field about 
423\,s after the trigger. 

To obtain an arcsecond level position of this source, 
we used the first 2775\,s PC mode data and, by correcting for the astrometric 
errors by utilizing Swift/UVOT data, 
we found 
    RA(J$2000)=18^{\rm h} 45^{\rm m} 01\fs58$, 
    Dec(J$2000)= -04^{\circ} 33^{\prime}  57\farcs4$, 
with an estimated error of 1\farcs4 radius (90\,\% c.l.). 
The XRT position is $2\farcs2$ from  the most likely optical counterpart proposed by 
\citet[][$\sim 1^{\prime\prime}$ error]{Coe1996:18450},
and it is the best currently available X--ray position for this source since
\citet[][ based on {\it XMM-Newton}]{Zurita2009:1845}  reports a localization  with an error of $2^{\prime\prime}$.

We extracted the mean BAT mask-weighted spectrum  (T$-$239 to T$+$963\,s; 1202\,s net exposure) 
during the first orbit of data and fit it in the 15--70\,keV energy band with a simple
power law. We obtained 
$\Gamma_{\rm BAT}=2.4\pm0.4$ ($\chi^2_{\nu}=0.9$, 24  dof), 
and a 15--70\,keV flux of  $1.7\times10^{-9}$ erg cm$^{-2}$ s$^{-1}$.

\begin{figure}[t]
\begin{center}
\vspace{+0.4truecm}
\centerline{\includegraphics[angle=270,width=8.5cm]{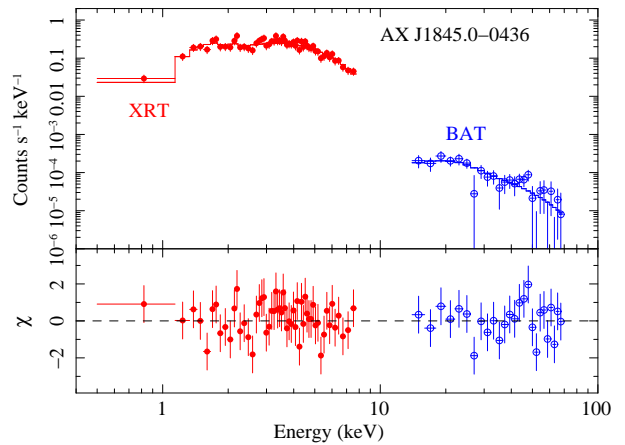}}
\end{center}
\caption{Broad-band spectroscopy of the 2012 May 05 outburst of AX~J1845.0$-$0433.    
		Top: XRT/PC (filled red circles) and BAT (empty blue circles) data 
			fit with the {\tt cutoffpl} model. 
		Bottom: the residuals of the fit (in units of standard deviations). }
\label{cospar12:fig:18450meanspec}
\end{figure}

Figure~\ref{cospar12:fig:best_sfxts} (panel h) shows the XRT light curve that we followed for a total of 18 days. 
In the XRT AX J1845.0 $-$0433 reached an initial peak at about 20 count s$^{-1}$ at T$+$1000\,s, immediately 
followed by several more flares. 
The following day, the source was not detected (3$\sigma$ upper limit at about $2.4\times10^{-2}$ count s$^{-1}$), 
and in the following days, it showed several more flares. 
The overall dynamic range therefore reaches at least 750 which places AX~J1845.0$-$0436 
neatly in the typical SFXT range. 
The initial XRT/WT spectrum (T$+$429 to T$+$846\,s; 417\,s net exposure) can be fit with 
an absorbed power law, with a photon index of $\Gamma=1.3\pm0.1$, and an absorbing column density 
of $N_{\rm H}=(1.8_{-0.2}^{+0.3}\times 10^{22}$ cm$^{-2}$,  in slight excess of the Galactic value 
of $1.30\times 10^{22}$ cm$^{-2}$,  \citealt{LABS}). 
The mean flux is $F_{\rm 2-10\,keV}=7\times10^{-10}$ erg cm$^{-2}$ s$^{-1}$ (unabsorbed). 
The XRT/PC spectrum (T$+$848 to T$+$1891\,s; 1043\,s net exposure) 
shows a power law shape with $N_{\rm H}=(1.4_{-0.3}^{+0.4})\times 10^{22}$ cm$^{-2}$, 
$\Gamma=0.8\pm0.2$ 
and an average flux of $F_{\rm 2-10\,keV}=8\times10^{-10}$ erg cm$^{-2}$ s$^{-1}$  (unabsorbed). 
These fluxes translate into a luminosity of $L_{\rm 2-10\,keV} = 1\times10^{36}$ erg s$^{-1}$ 
\citep[assuming the optical counterpart distance of 3.6\,kpc, ][]{Coe1996:18450}, comparable 
with the one observed in previous outbursts of this source \citep[][]{Sguera2007:18450,Zurita2009:1845}. 

We fit the nearly simultaneous BAT and XRT/PC spectra 
in the 0.5--10\,keV and 15--70\,keV energy bands for XRT and BAT, respectively,
with the same models as those used for \src. 
Our results are reported in Table~\ref{cospar12:tab:18450broadspec}. 
The simple power-law model yields a clearly unsatisfactory fit, showing systematic residuals
that indicate the need for a curved spectrum. 
The absorbed power-law with a high energy cut-off yields a poorly constrained
value of the cut-off energy. The absorbed power-law with an exponential cut-off,
on the other hand, yields a better fit 
(see Fig.~\ref{cospar12:fig:18450meanspec}),
with $N_{\rm H}=(1.3_{-0.3}^{+0.4})\times 10^{22}$ cm$^{-2}$,  $\Gamma=0.56_{-0.25}^{+0.26}$,
and a high-energy cut-off at $E_{\rm c}=17_{-5}^{+8}$ keV. 
These results are consistent with those of \citet[][]{Sguera2007:18450} 
and \citet[][]{Zurita2009:1845} both in terms of overall luminosity of the outbursts 
and detailed spectral parameters.

\section{Conclusions and Future Perspectives}

In this paper we have given a review of our {\it Swift} SFXT Project, its underlying observing 
strategy and its advantages. 
Given the shape of the SFXT spectrum,  broad band spectroscopy with {\it Swift} (0.3--150\,keV)  
generally allows us to both model the hard-X spectral properties 
and to measure the absorption. Furthermore, we can use our detailed light curves to determine the 
overall dynamic range, a discriminant between outbursts of classical sgHMXBs and SFXTs.

We used two recent outbursts to highlight the typical results of our activity. 
\src\ is considered a candidate SFXT, based on its recorded history of flaring activity
and lack of a confirmed counterpart.
Our evidence, based on \sw\ data, points towards an intermediate SFXT classification. 
The temporal and spectroscopic properties of this source, 
including the high peak luminosities, the large dynamic range, and the length (days) and flaring 
nature of the X--ray outburst are strongly reminiscent of those of the prototypes of the SFXT class. 
AX~J1845.0$-$0436, on the other hand, is a confirmed SFXT with a well-known optical companion.
Our \sw\  simultaneous broad band spectroscopy allows us to obtain a good fit with 
an absorbed flat power-law with an exponential cut-off at $\sim 18$\,keV, 
as is typical for wind-fed HMXBs.
In Fig.~\ref{cospar12:fig:best_sfxts} we draw a comparison among the best light curves  
of outbursts of confirmed SFXTs and candidate SFXTs, as caught during 
our monitoring campaigns with \sw. Similarly to the observations for 
all confirmed SFXTs, the outburst of \src\  (hence, in the SFXT framework, 
the length of the accretion phase) lasts a few days. 
In particular, we show for the first time, a long monitoring of AX~J1845.0$-$0436
after an outburst, also showing remarkable activity for several days, at least.  
Furthermore, in both cases we also observed 
the multiple-peak structure of the light curve that can now be considered a defining 
characteristic of the SFXT class, which is likely due to inhomogeneities in the accretion flow
\citep[e.g.][]{zand2005} 
and/or unstable accretion due to the spin and magnetic field of the compact object
\citep[][]{Grebenev2007,Bozzo2008}. 

Currently the Project is catching bright flares at the rate of several per year,
that we follow with XRT for at least one week (often longer if previously not observed in the soft X--rays) 
through ToO observations through the {\it Swift} GI program (PI: P.\ Romano) and regular ToOs. 
We aim at observing bright flares of the remaining candidate SFXTs in order to localize them to the arcsecond level and
identify their optical/infrared companion,
as well as the remaining SFXTs for which no soft X--ray observations are
available, yet. 
At the time of writing long term monitoring campaigns are still ongoing to systematically
study the out-of-outburst behaviour of the whole SFXT class.

\section{Acknowledgements}

We thank the \sw\ team duty scientists and science planners 
and the remainder of the \sw\ XRT and BAT teams, 
S.\ Barthelmy in particular, for their invaluable help and support. 
We also thank D.~Grupe and M.M.~Chester for helpful discussions. 
{We also thank the anonymous referees for their insightful comments. 
We acknowledge financial contribution from the agreement ASI-INAF I/009/10/0 
and from contract ASI-INAF I/004/11/0.  
This work made use of the results of the Swift/BAT hard X-ray transient monitor: \\
\href{http://swift.gsfc.nasa.gov/docs/swift/results/transients/}{http://swift.gsfc.nasa.gov/docs/swift/results/transients/} \\
and of data supplied by the UK Swift Science Data Centre at the University of Leicester.

\end{document}